\documentclass[twocolumn,showpacs,preprintnumbers,amsmath,amssymb]{revtex4}

\usepackage{graphicx}
\usepackage{dcolumn}
\usepackage{bm}

\begin{document}

\preprint{}

\title{Imaging spin flows in semiconductors subject to electric, magnetic, and strain fields}

\author{S.~A.~Crooker}
\author{D.~L.~Smith}

\affiliation{Los Alamos National Laboratory, Los Alamos, NM 87545}

\date{November 13, 2004}

\begin{abstract}
Using scanning Kerr microscopy, we directly acquire two-dimensional images of spin-polarized
electrons flowing laterally in bulk epilayers of n:GaAs.  Optical injection provides a local dc
source of polarized electrons, whose subsequent drift and/or diffusion is controlled with electric,
magnetic, and - in particular - strain fields. Spin precession induced by controlled uniaxial
stress along the $\langle$110$\rangle$ axes demonstrates the direct {\bf k}-linear spin-orbit
coupling of electron spin to the shear (off-diagonal) components of the strain tensor,
$\epsilon_{xy}$.
\end{abstract}


\maketitle

The ability to control and measure electron spin degrees of freedom in semiconductors has been
proposed as the operating principle for a new generation of novel electrical devices with the
potential to overcome the power consumption and speed limitations of conventional electronic
circuits \cite{Wolf,Zutic}. Semiconductor devices utilizing electron spin generally require: i)
transport of spin-polarized electrons from one location in the device to another, and ii) a means
to manipulate the electron spin orientation, either directly with magnetic fields or indirectly
with electric and/or strain fields that exploit the spin-orbit interaction by coupling to electron
orbital motion. Many spin-based semiconductor device proposals
\cite{Datta,Schliemann,Hall,Cartoixa} are based on a field-effect transistor geometry in which
electron transport occurs in essentially 2-dimensional structures.  In order to design
semiconductor structures whose function is based on electron spin it is necessary to understand the
transport and flow of spin-polarized electrons, and how it is influenced by electric, magnetic and
strain fields in these 2D structures.

Using methods for scanning Kerr microscopy, we acquire 2D images of spin-polarized conduction
electrons flowing laterally in bulk epilayers of n-type GaAs.  The images directly reveal the
spatial dependence of spin diffusion and spin drift in the presence of applied electric, magnetic,
and - in particular - strain fields. Controlled uniaxial stress along the $\langle$110$\rangle$
axes induces spin precession, revealing the direct ({\bf k}-linear) spin-orbit coupling of
electron spin to the off-diagonal components of the strain tensor {\boldmath $\epsilon$}. The
coupling may be characterized by an effective strain-induced magnetic field $\mathbf{B}_\epsilon$,
which is shown to be orthogonal to the electron momentum \textbf{k}, and therefore chiral for
radially-diffusing spins. $\mathbf{B}_\epsilon$ scales linearly with $|\mathbf{k}|$, yielding a
spatial precession of electron spins that is independent of electrical bias and is considerably
more robust against the randomizing (ensemble dephasing) effects of spin diffusion as compared
with precession induced by external magnetic fields.

The samples are 1 $\mu$m thick, silicon-doped (n-type) GaAs epilayers grown by molecular beam
epitaxy on [001]-oriented semi-insulating GaAs substrates. Doping densities are n$_e$= 1, 5, and
10$\times$10$^{16}$/cm$^3$.  Pieces were cleaved along the [110] and ${[1\bar{1}0]}$ natural cleave
directions, and ohmic contacts allowed an in-plane, lateral electrical bias in the [110] direction.
The samples were mounted in vacuum on the cold finger of an optical cryostat. To apply controlled
uniaxial stress along the [110] or ${[1\bar{1}0]}$ axis at low temperature, the cold finger
incorporated a small cryogenic vise, whose lead screw was adjusted via a retractable actuator.
Since all the samples exhibited qualitatively similar strain-related effects, we show data from
only the n$_e$=10$^{16}$/cm$^3$ epilayer.

A local, steady-state source of electrons, spin polarized along the sample normal ([001], or
${\hat{z}}$), was provided by a circularly-polarized 1.58 eV laser focused to a 4 $\mu$m spot on
the epilayer. While this pump laser (typically 10-25 $\mu$W) injects 50\% spin-polarized electrons
{\it and} holes, the holes spin-relax and recombine rapidly, leaving behind a net spin polarization
of the mobile conduction electrons \cite{Kikkawa}. These polarized electrons subsequently drift
and/or diffuse laterally away from the point of generation.  2D images of the resulting {\it
z}-component of electron spin polarization were acquired by measuring the polarization (Kerr)
rotation imparted on a linearly-polarized probe laser that was reflected from the epilayer surface
and raster-scanned in the {\it x-y} epilayer plane.  The probe beam (50-100 $\mu$W) was derived
from a narrowband tunable cw Ti:sapphire laser, and also focused to a 4 $\mu$m spot. For lock-in
measurement, the pump laser polarization was modulated from left- to right-circular (injecting
spins along ${\pm \hat{z}}$) at 51 kHz.

The measured Kerr rotation is a strong function of probe laser energy near the GaAs bandgap
(inset, Fig. 1a).  Here, 30 $\mu$m separates the pump and probe spots, so that the signal arises
solely from a non-zero spin polarization of the electron Fermi sea. This energy-dependent response
provides a relative (and \textit{in-situ}) monitor of stress-induced bandedge shifts. When
imaging, the probe energy is tuned (as shown) to the broad maximum that exists \textit{below} the
1.515 eV GaAs bandedge, to avoid perturbation of the Fermi sea.

\begin{figure}[tbp]
\includegraphics[width=.33\textwidth]{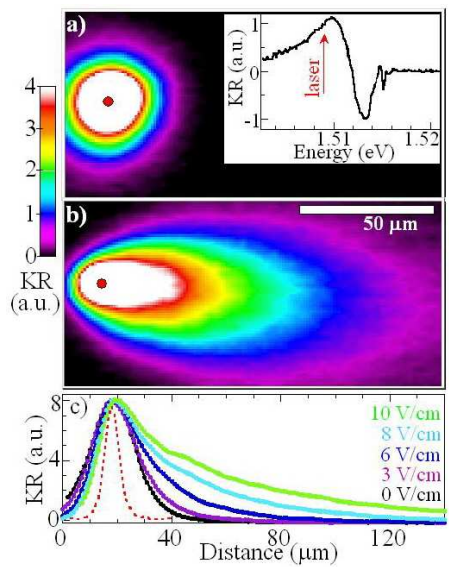}
\caption{  (a) 70$\times$140 $\mu$m image of electron spin polarization in a 1 $\mu$m thick n:GaAs
epilayer (n$_e$=10$^{16}$cm$^{-3}$) at 4K, acquired via Kerr-rotation (KR) microscopy. A
circularly-polarized 1.58 eV laser focused to a 4 $\mu$m spot provides a local, dc source of spin
polarized electrons; this image indicates 2D spin diffusion. Inset: KR vs. probe energy near 1.515
eV GaAs bandedge. (b) With E=10 V/cm lateral electrical bias, showing spin diffusion and drift. (c)
Cross-sections of spin flow vs. bias; dotted line shows 5.5 $\mu$m resolution.} \label{fig1}
\end{figure}

Fig. 1a shows a 70$\times$140 $\mu$m image of the steady-state electron spin polarization in the
n$_e$=10$^{16}$/cm$^3$ GaAs epilayer due to spin diffusion alone.  To emphasize smaller signals,
the color scale has been adjusted so that white equals half the peak signal. The spatial extent of
the measured spin polarization ($\sim$60 $\mu$m edge to edge) is much larger than the focused pump
laser (shown by the red spot), indicating diffusion of electron spins away from the point of
generation. Images of radially-diffusing electrons provide information on spin flows along {\it
all} {\bf k} directions in the {\it x-y} sample plane, which will prove useful later in confirming
the direction of $\mathbf{B}_\epsilon$. The characteristic diffusion length is given by the
electron spin lifetime $\tau_s$ (measured independently via ultrafast techniques) and spin
diffusion constant $D_s$. Fits to a 2D drift-diffusion model (described later) indicate
$D_s\simeq$3 and 15 cm$^2$/s for the n$_e$=1 and 5$\times$10$^{16}$/cm$^3$ samples at 4K
respectively, values in accord with the charge diffusion constants. Spin-polarized electrons can
also be induced to drift laterally along an applied electric field E (Fig. 1b). This ``spin drag",
as revealed by Kikkawa and Awschalom \cite{Kikkawa}, occurs over length scales $>$100 $\mu$m.  2D
images of spin drift and diffusion (Fig. 1b, where E=10 V/cm) show roughly elliptical contours of
constant spin polarization, with major axis determined by electrical bias and lateral spin flow
extending beyond 150 $\mu$m. Fig. 1c shows normalized line cuts through a series of images.

Spin-orbit coupling in GaAs permits coupling to electron spin degrees of freedom through the
\textit{spatial} part of the electron wavefunction, thus allowing induced precession of electron
spins \textit{without} external magnetic fields \cite{Kato,Flatte,Winkler}.  Spin-orbit effects
lead to spin splittings of the conduction band along particular crystal momenta {\bf k}, and can be
characterized by effective magnetic fields. Bulk inversion asymmetry (BIA) arises from the lack of
inversion symmetry in GaAs, leading to a spin-splitting for electrons with {\bf k} along the
$\langle$110$\rangle$ axes (but no splitting along $\langle$111$\rangle$ or $\langle$100$\rangle$
axes). This ubiquitous coupling, cubic in $|\mathbf{k}|$, is the origin of the D'yakonov-Perel'
mechanism of electron spin relaxation in bulk GaAs.  A second spin-orbit effect arises from
structural inversion asymmetry (the SIA or ``Rashba" term \cite{Rashba}), as typically found in 2D
heterostructures. Inversion asymmetry of the confining potential along the growth direction
$\hat{z}$ (typically [001]) can often be characterized by an electric field $\mathbf{E}_z$, giving
a Rashba Hamiltonian $H_R \propto$ {\boldmath $\sigma$} $\cdot(\mathbf{k}\times\mathbf{E}_z)$. For
typical 2D heterostructures with {\bf k} in the {\it x-y} plane, the effective magnetic field
$\mathbf{k}\times\mathbf{E}_z$ is therefore in-plane and orthogonal to {\bf k}, with magnitude
linear in $|\mathbf{k}|$. Control of the Rashba spin-orbit term, through a gate-tunable
$\mathbf{E}_z$, is the basis of the original Datta-Das spin transistor \cite{Datta}.

\begin{figure}[tbp]
\includegraphics[width=.33\textwidth]{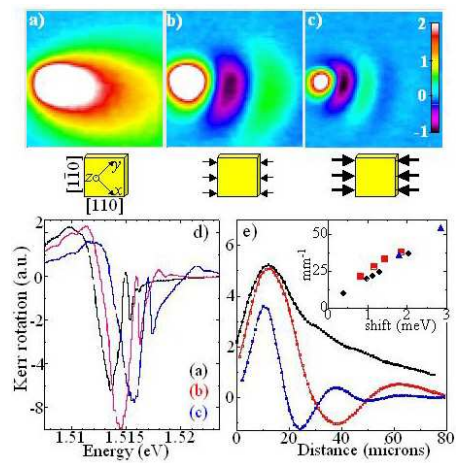}
\caption{(a-c) 80$\times$80 $\mu$m images of 2D spin flow (E=10 V/cm) at 4K, showing induced spin
precession with increasing [110] uniaxial stress. (d) KR vs. probe energy for the images, showing
blueshift of GaAs bandedge with stress. (e) Line cuts through the images. Inset: Spatial frequency
of spin precession vs. bandedge shift.} \label{fig2}
\end{figure}

Figure 2 demonstrates an additional spin-orbit effect; namely the coupling of the electron spin to
the strain tensor {\boldmath $\epsilon$}. As described previously \cite{Pikus,Seiler,Cardona,Bir},
stress along the $\langle$110$\rangle$ axes of GaAs induces {\bf k}-linear spin splittings in the
conduction band through the off-diagonal (shear) elements of {\boldmath $\epsilon$}. The strain
Hamiltonian is $H_S=c_3${\boldmath $\sigma\cdot\varphi$}, where
$(\varphi_x,\varphi_y,\varphi_z)=(\epsilon_{xy}k_y-\epsilon_{xz}k_z,
\epsilon_{yz}k_z-\epsilon_{yx}k_x, \epsilon_{zx}k_x-\epsilon_{zy}k_y)$, $(x,y,z)$ are the principle
$\langle$100$\rangle$ crystal axes, and the constant $c_3$ depends on the interband deformation
potentials. Stress applied along the [110] or ${[1\bar{1}0]}$ axis of GaAs gives in-plane shear
$\epsilon_{xy}=\epsilon_{yx}\neq0$ . Thus for electrons moving in the {\it x-y} plane, $H_S$ has
similar symmetry to $H_R$, as it describes an in-plane effective magnetic field, orthogonal to {\bf
k}, with magnitude linear in $|\mathbf{k}|$.

Figs. 2a-c show 80$\times$80 $\mu$m images of steady-state electron spin flow
($\mathbf{k}\parallel$ [110]) in the presence of increasing [110] uniaxial stress.  Spin precession
is observed, with increasing spatial frequency, indicating a strain-induced effective magnetic
field $\mathbf{B}_{\epsilon}\propto\epsilon_{xy}|\mathbf{k}|$, oriented along ${[1\bar{1}0]}$.
$\epsilon_{xy}$ is inferred from the measured blueshifts (see Fig. 2d) of $\sim$1 and 2 meV in
Figs. 2b and 2c respectively, indicating strain $|\epsilon_{xy}|\sim$1.5 and 3.0$\times$10$^{-4}$
(and applied stress $\sim$3.6 and 7.2$\times$10$^8$ dynes/cm$^{-2}$ \cite{Bir}). These strains are
small compared to typical $\sim$1\% strains due to lattice mismatched growth; in fact, considerable
care was required in sample mounting to avoid spurious and inhomogeneous strains during cool-down.
With the cryogenic vise, the stress-induced precession of electron spins is controllable,
reversible, and uniform over the sample. Line cuts along [110] (Fig. 2e) show many precession
cycles ($>5\pi$ rotation). The inset of Fig. 2e confirms that the spatial frequency of the induced
precession ($\propto\mathbf{B}_\epsilon$) scales linearly with the observed bandshift ($\propto
\epsilon_{xy}$).

\begin{figure}[tbp]
\includegraphics[width=.33\textwidth]{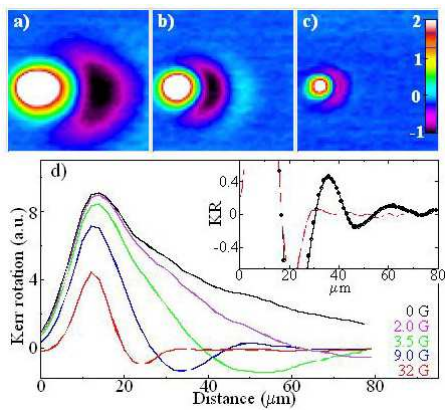}
\caption{(a-c) 80$\times$80 $\mu$m images of 2D spin flow (E=10 V/cm) at 4K, with increasing
applied magnetic field $\mathbf{B}_{app}$=3.5, 9, and 32 G along ${[1\bar{1}0]}$. (d) Line-cuts
through the images. Inset: Comparing line-cuts through Figs. 2c (black) and 3c (red).}
\label{fig3}
\end{figure}

Only shear (off-diagonal) strain gives \textbf{k}-linear spin-orbit coupling to electron spins.
Strain along the principle axes, either applied or arising from, {\it e.g.}, lattice mismatched
growth along [001], should not influence electron spins to leading order. However, shear strain and
\textbf{k}-linear coupling \textit{should} exist in lattice-mismatched structures grown along [110]
or [111]. Because shear strain reduces the symmetry of the zincblende GaAs crystal, these strain
effects can be thought of as arising from BIA in a crystal of lower symmetry. In lower symmetry
wurtzite crystals such as CdSe or ZnO, BIA alone gives \textbf{k}-linear spin splitting of
electrons \cite{Voon}. One should not think of $\mathbf{B}_\epsilon$ as arising from an electric
field (\textit{e.g.}, a stress-induced piezoelectric field \cite{Fricke}), since any such field is
screened in bulk metallic samples.

The line-cuts in Fig. 2e show that dc spin flows, precessing due to strain, are in phase at large
distances from the point of generation. This robust behavior is in marked contrast with the rapid
spatial dephasing that occurs when external magnetic fields are used to induce precession in dc
spin flows, as shown in Fig. 3  Figs. 3 a-c show spin flows in the presence of an increasing
applied magnetic field ($\mathbf{B}_{app}\parallel [1\bar{1}0]$), with line-cuts in Fig. 3d. As
$\mathbf{B}_{app}$ increases, the ensemble spin polarization becomes dephased at distances beyond
one precession period, particularly when the precession period falls below the spin diffusion
length. This pronounced spatial dephasing of dc spin flows is due to the randomizing nature of
diffusion.  The net spin at a remote location is the combined sum of many random walks. Each path
takes a different amount of time, giving a different degree of spin precession, leading to
dephasing. Future spintronic devices based on magnetic field manipulation of diffusive spin flow
may thus be practically limited to a regime requiring $\pi$ rotation or less.

\begin{figure}[tbp]
\includegraphics[width=.33\textwidth]{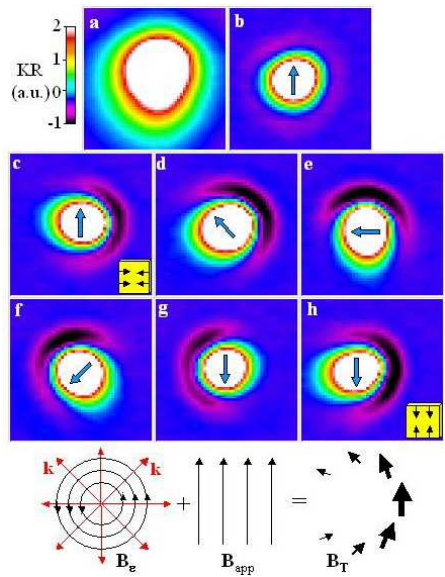}
\caption{50$\times$50 $\mu$m images of 2D spin diffusion (E=0) at 4K.  (a) Stress=0,
$\mathbf{B}_{app}$=0. (b) Stress=0, $\mathbf{B}_{app}$=16 G oriented along ${[1\bar{1}0]}$ as
shown. (c) $\mathbf{B}_{app}$=16 G, with [110] uniaxial stress.  Spins diffusing to the right
precess, while those diffusing to the left do not. Thus $\mathbf{B}_\epsilon$ is chiral for
radially-diffusing electrons (see diagram). (d-g) Maintaining [110] stress, $\mathbf{B}_{app}$ is
rotated by 180 degrees in-plane. (h) Stress is switched to ${[1\bar{1}0]}$, reversing chirality of
$\mathbf{B}_\epsilon$.} \label{fig4}
\end{figure}

A comparison of linecuts through Figs. 2c and 3c are shown in the inset of Fig. 3d. Clearly, the
spatial coherence of dc spin flows is preserved over more precession cycles (and greater distance)
when the spins are manipulated with strain instead of magnetic field. This is a direct consequence
of the $|\mathbf{k}|$-linear nature of $\mathbf{B}_\epsilon$, which correlates precession
frequency with electron velocity (and therefore position). Indeed, if electrons moved along only
one dimension, the spin flow would not dephase at all (it would still, of course, decohere).
Scattering from +\textbf{k} to -\textbf{k} would simply reverse the direction of precession,
leading to an exact correspondence between spatial position and spin orientation.

The images in Fig. 4 confirm that $\mathbf{B}_\epsilon$ is orthogonal to \textbf{k}, such that
$\mathbf{B}_\epsilon$ circulates around the point of injection for radially diffusing electrons.
Image (a) shows unperturbed spin diffusion (no stress and $\mathbf{B}_{app}$=0). In (b),
$\mathbf{B}_{app}$=16 G along ${[1\bar{1}0]}$ as indicated. Spins precess uniformly, regardless of
\textbf{k}, giving a faint annulus of oppositely oriented spins (negative signal) surrounding the
injection point. In (c), stress is applied along [110], and the image becomes asymmetric. Electrons
diffusing to the right along $\mathbf{k}\parallel$ [110] undergo precession, while those diffusing
to the left along -\textbf{k} do not.  That is, the total field
($\mathbf{B}_T$=$\mathbf{B}_\epsilon$+$\mathbf{B}_{app}$) is finite for spins diffusing to the
right, but $\mathbf{B}_T$ is effectively zero for spins diffusing to the left. This image is
consistent with a uniform $\mathbf{B}_{app}$ added to a circulating $\mathbf{B}_\epsilon$, as
shown. Maintaining [110] stress, the image asymmetry rotates and ultimately reverses as
$\mathbf{B}_{app}$ is rotated in the {\it x-y} plane through 180 degrees (c-g). Finally (h), when
stress is applied along the ${[1\bar{1}0]}$ axis, the asymmetry again reverses, indicating the
opposite chirality of $\mathbf{B}_\epsilon$.

\begin{figure}[tbp]
\includegraphics[width=.33\textwidth]{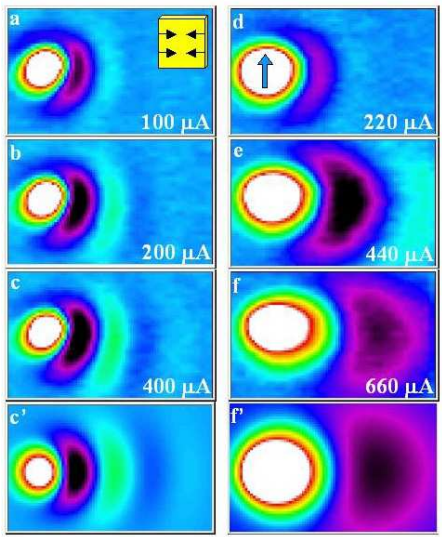}
\caption{50$\times$80 $\mu$m images of 2D spin flow at 4K with increasing $|\mathbf{k}|$ (current).
(a-c) With [110] stress and $\mathbf{B}_{app}$=0. Spatial period of precession is fixed. (d-f)
Stress=0, $\mathbf{B}_{app}$=6 G; spatial period varies. Images c') and f') are simulations of c)
and f), using the model described in text.} \label{fig5}
\end{figure}

For comparison with data, we derive and solve the spin drift-diffusion equations that describe dc
spin flows in the presence of electric, magnetic and strain fields. For simplicity, the [110]
strain axis is taken here to be the {\it x}-axis, and the electric and magnetic fields are in the
{\it x-y} sample plane.  The spin polarization is described by the ($\rho_x$, $\rho_y$, $\rho_z$)
components of a 2$\times$2 density matrix, where $\rho_z$ gives the ensemble spin density. The
equations are $O_1\rho_x=-O_2\rho_z$, $O_1\rho_y=-O_3\rho_z$, and
$O_4\rho_z-O_2\rho_x-O_3\rho_y=-G_z$, with operators $O_1 = D\nabla ^2  + \mu \mathbf{E}\cdot
\mathbf{\nabla} - (C_s \varepsilon)^2 D - 1/T_2 $, $O_2  = - C_B B_y  + C_s \varepsilon (2D\nabla
_x + \mu E_x )$, $O_3  = C_B B_x  + C_s \varepsilon (2D\nabla_y  + \mu E_y )$, and $O_4 = D\nabla
^2 + \mu \mathbf{E} \cdot \mathbf{\nabla} - 2(C_S \varepsilon)^2 D - 1/T_1$. $D$ is the spin
diffusion constant, $\mu$ is the electron mobility, $T_2$ and $T_1$ are transverse and
longitudinal spin lifetimes, the magnetic coupling constant is $C_B=\frac{g\mu_B}{\hbar}$, the
strain coupling constant $C_s$ is given in Ref.\cite{Bir}, $G_z$ is the spin generation rate, and
$\varepsilon$ is the off-diagonal strain element $\epsilon_{xy}$. Drift and diffusion appear in
the strain coupling terms, due to its dependence on $\mathbf{k}$. The 1 $\mu$m thick n:GaAs layer
is small compared with a lateral spin diffusion length, thus, the problem is 2-dimensional.  The
equations are solved with numerical Fourier transform methods.

In contrast with the case of an applied magnetic field, both the model and the data reveal a
consequence of {\bf k}-linear $\mathbf{B}_\epsilon$: the spatial period of precession is {\it
independent} of applied electrical bias. Fig. 5 confirms this, showing spin flow along [110]. In
the presence of strain (Fig. 5a-c), the spatial precession period is independent of electrical
bias, whereas for the case of applied magnetic field (d-f), the spatial precession period clearly
increases with increasing electrical bias. Functional devices based on rotation of spin from one
point in space to another ({\it e.g.}, from source to drain contacts in spin transistor designs
\cite{Datta,Schliemann,Hall,Cartoixa}) may well benefit from the freedom to operate at variable
electrical bias. Along with spin manipulation via Rashba coupling (also linear in $|\mathbf{k}|$),
the spin-orbit coupling of spin flows to shear strain, as detailed in this work, also affords this
flexibility.

This work was supported by the DARPA SpinS and the Los Alamos LDRD programs.  We thank S. Kos, P.
Littlewood, and I. Martin for valuable discussions.


\end{document}